\documentstyle[11pt,aaspp4,psfig]{article}
\input psfig
\psfull
\singlespace

\parskip=\the\medskipamount

\def\h{ {\it h} }
\def\om{\Omega_{m}}

\def\omzero{\Omega^{\Lambda=0}_{m}}
\def\ol{\Omega_{\Lambda}}

\def\obh{\Omega_{b} \h^{2}}
\def\t{ t_{o}}
\def\tg{ t_{Gal}}
\def\td{t_{disk}}
\def\tf{\t^{flat}}
\def\etal{{\it et al.}\/}
\def\etall{{\it {\sf et al.} \/}}
\def\hg{{\it {\mathsf h}}}
\def\omg{{\mathsf \Omega_{m}}}
\def\omfg{{\mathsf \Omega^{flat}_{m}}}
\def\omzerog{{\mathsf \Omega^{\Lambda=0}_{m}}}
\def\olg{{\mathsf \Omega_{\Lambda}}}
\def\obg{{\mathsf \Omega_{b}}}
\def\obhg{{\mathsf \Omega_{b} \hg^{2}}}
\def\tog{{\it {\mathsf t_{o}}}}
\def\tgg{{\it {\mathsf t_{Gal}}}}
\def\tdg{{\it {\mathsf t_{disk}}}}
\def\tfg{{\it \t^{flat}}}

\def\etallg{{\it {\sf et al.} \/}}

\newcommand {\lsim}{\mbox{$\:\stackrel{<}{_{\sim}}\:$} }

\def\be{\begin{equation}}
\def\ee{\end{equation}}
\begin{document}
\raggedright
%\begin{sffamily}
\begin{center}
\title{\huge
{\bf {\sf A Younger Age for the Universe}}}
\medskip
\author{ {\Large {\bf {\sf  Charles H. Lineweaver}}}}
School of Physics, University of New South Wales, 
Sydney NSW 2052, Australia \\
charley@bat.phys.unsw.edu.au
\end{center}
%\newpage
\begin{abstract}
%{\large \sf 
The age of the universe in the  Big Bang model can be calculated from three parameters: 
Hubble's constant, $\hg$; the mass density of the universe, $\omg$; and the cosmological constant, $\olg$.
Recent observations of the cosmic microwave background and 
six other cosmological measurements reduce the uncertainty in these three parameters,
yielding an age for the universe of  $13.4 \pm 1.6$ billion years,
which is a billion years younger than other recent age estimates.
A different standard Big Bang model, which includes cold dark matter with a cosmological constant, 
provides a consistent and absolutely time-calibrated evolutionary sequence 
for the universe.
%}    %end \sf
\end{abstract}
%\end{sffamily}
%\newpage

In the Big Bang model, the age of the universe, $\t$,
is a function of three parameters: $\h$, $\om$ and $\ol$ ({\it 1}).
%{\bf If we know them, we know $\t$}.
The dimensionless Hubble constant, $\h$,
%= (H_{o}/100\; km\; s^{-1} Mpc^{-1})$,
tells us how fast the universe is expanding.
The density of matter in the universe, $\om$, slows the expansion, and the cosmological
constant, $\ol$,  speeds up the expansion (Fig. 1).

Until recently, large uncertainties in the measurements of $\h$, $\om$ and $\ol$ made efforts to determine 
$\t(\h,\om,\ol)$ unreliable.
Theoretical preferences were, and still are, often used to remedy these 
observational uncertainties. One assumed the standard model ($\om = 1$, $\ol = 0$),
dating the age of the universe to $\t = 6.52/h$ billion years old (Ga).
However, for large or even moderate $\h$ estimates ($ \gtrsim  0.65$),
these simplifying assumptions resulted in an age crisis in which the universe 
was younger than our Galaxy ($\t \approx 10$ Ga $< \tg \approx 12 $ Ga).
These assumptions  also resulted in a baryon crisis in which estimates of the 
amount of normal (baryonic) matter in the universe were in conflict ({\it 2, 3}).

Evidence in favor of $\om < 1$ has become more compelling ({\it 4-8}), but $\ol$ is still 
often assumed to be zero, not because it is measured to be so, 
but because models are simpler without it.
Recent evidence from supernovae (SNe) ({\it 4, 5}) 
%(Perlmutter \etal 1998, Riess \etal 1998). 
indicates that $\ol > 0$.
These SNe data and other data exclude the standard Einstein-deSitter model ($\om = 1$, $\ol =0$).
The cosmic microwave background (CMB), on the other hand, excludes models with low $\om$ and $\ol = 0$ ({\it 3}). 
%(Lineweaver \& Barbosa 1998). 
With both high and low $\om$ excluded, $\ol$ cannot be zero.
Combining CMB measurements with SNe and other data,
I ({\it 9}) have reported $\ol = 0.62 \pm 0.16$.
[see ({\it 10-12}) for similar results].
%See also White \etal (1998), Tegmark (1998) and Efstathiou \etal (1999).
If $\ol \ne 0$, then estimates of the age of the universe in Big Bang models must 
include $\ol$. Thus one must use the most general form: $\t = f(\om,\ol)/h$  ({\it 13}). 
%(e.g. Felten \& Isaacman 1987, Carroll, Press and Turner 1992).

%%%%%%%%%%%%%%%%%%%%%%%%%%%%%%%%%%%%%%%%%%%%%%%%%%%%%%%%%%%%%%%%%%%%%%%%%%%%%%%%%%%%%%%%
\clearpage

%\vspace{15cm}
\begin{figure*}[!b]
\centerline{\psfig{figure=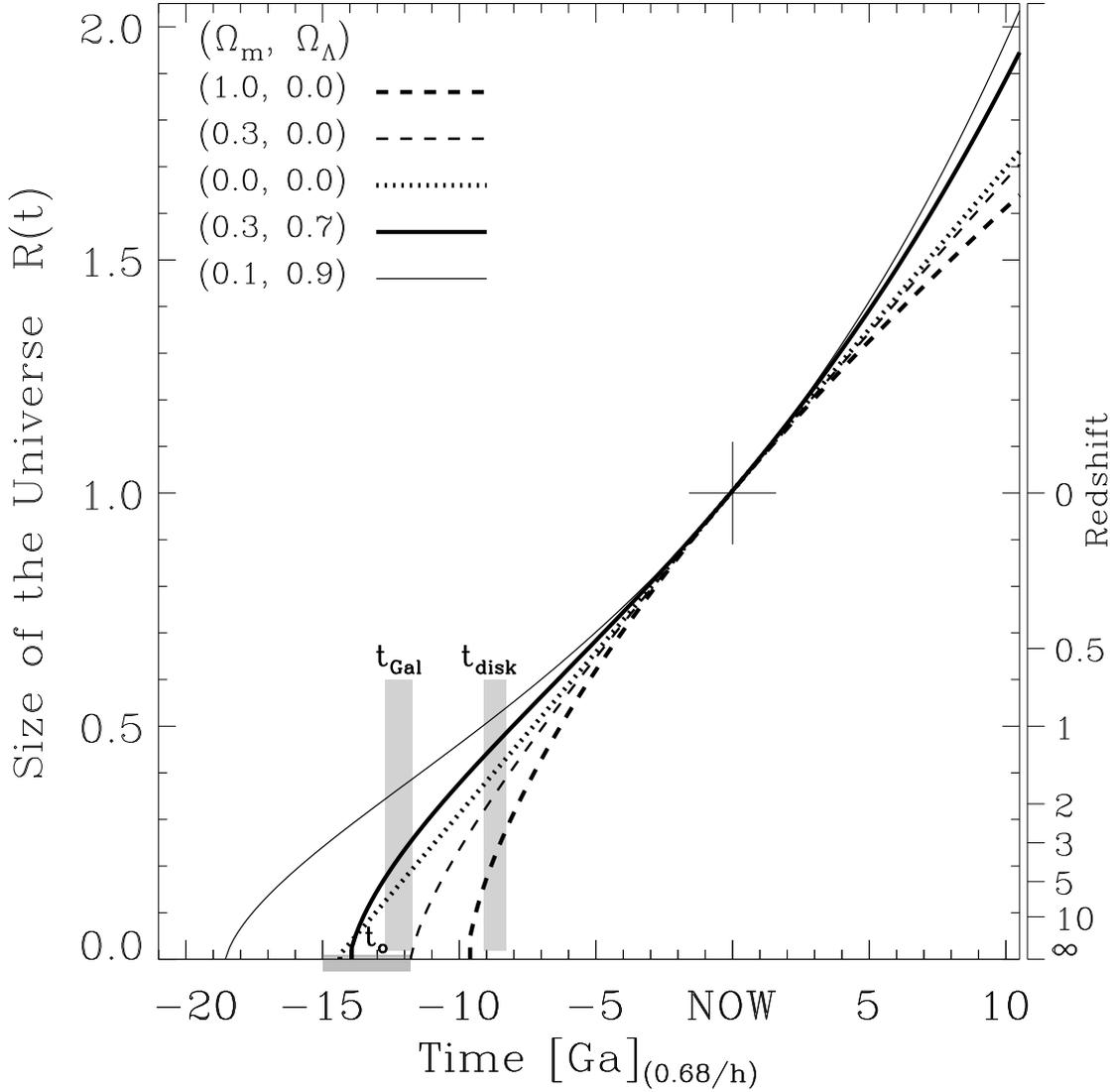,height=16.0cm,width=16.cm}}
%\centerline{\psfig{figure=~/lambda/figures/plotchi_triangle.ps,height=18.0cm,width=18.cm,bbllx=10pt,bblly=110pt,bburx=594pt,bbury=690pt}}
%BoundingBox: 28 70 594 636
\caption{The size of the universe, in units of its current size, as a function of time.
The age of the five models can be read from the x axis
as the time between NOW and the intersection of the
model with the x axis.
The main result of this paper, $\tog = 13.4 \pm 1.6$ Ga, is labeled ``$\tog$'' and is shaded 
gray on the x axis.
Measurements of the age of the halo of our Galaxy yield $\tgg = 12.2 \pm 0.5$ Ga, 
whereas measurements of the age of the disk of our Galaxy yield $\tdg = 8.7 \pm 0.4$ Ga
(Table 2). These age ranges are also labeled and shaded gray.
The  $({\mathsf \om, \ol)=(0.3,0.7)}$ model fits the constraints of Table 1  better than the other models
shown.
Over the past few billion years and on into the future, 
the rate of expansion of this model increases ($\ddot{R} > 0$). 
This acceleration means we are in a period of slow inflation.
%
%Not only the age but also the destiny of the Universe is determined by the estimates
%of $\hg$, $\omg$ and $\olg$. 
Other  consequences of a $\ol$-dominated universe are discussed in ({\it 50}).
On the x axis $\hg = 0.68$ has been assumed.
For other values of $\hg$, multiply the x axis ages by $0.68/\hg$.
Redshifts are indicated on the right.}
\label{fig:figon}
\end{figure*}
%%%%%%%%%%%%%%%%%%%%%%%%%%%%%%%%%%%%%%%%%%%%%%%%%

\clearpage

%\vspace{15cm}
\begin{figure*}[!b]
\centerline{\psfig{figure=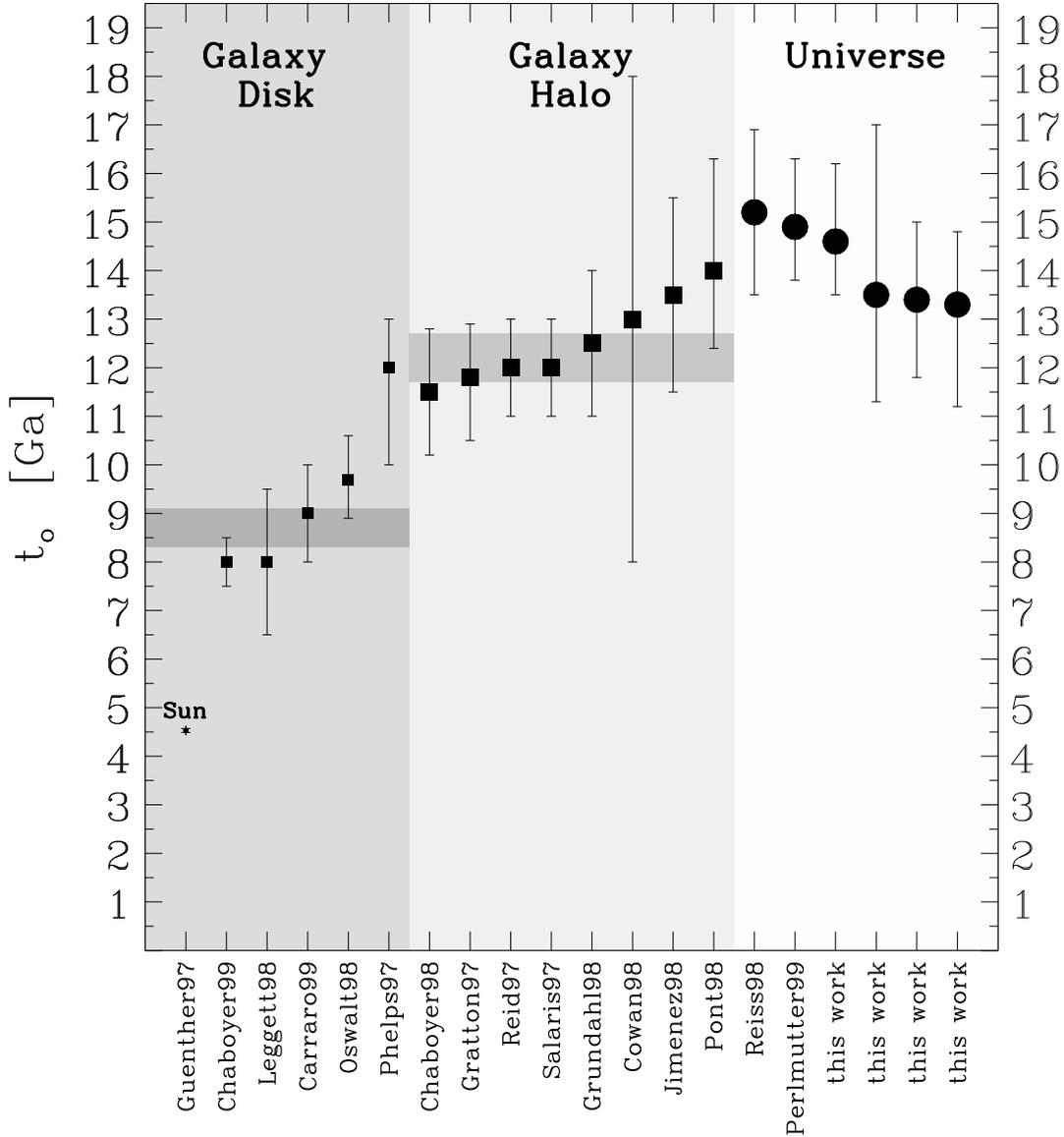,height=16.0cm,width=16.cm}}
%\centerline{\psfig{figure=~/lambda/figures/plotchi_triangle.ps,height=18.0cm,width=18.cm,bbllx=10pt,bblly=110pt,bburx=594pt,bbury=690pt}}
%BoundingBox: 28 70 594 636
\caption{Age estimates of the universe and of the oldest objects in our Galaxy.
The four estimates of the age of the universe from this work are indicated 
in Table 2.
The three similar points near $13.4$ Ga, result from $\hg = 0.64, 0.68, 0.72$
and indicate that the result is not strongly dependent on $\hg$ when
a reasonable $\hg$ uncertainty of $\pm 0.10$ is used.
Among the four, the highest value at 14.6 Ga comes from assuming $\hg = 0.64 \pm 0.02$.
All the estimates in the top section of Table 2 are plotted here.
As in Fig. 1, averages of the ages of the Galactic halo and Galactic
disk are shaded gray.
The absence of any single age estimate more than $\sim 2 \sigma$ from the average
adds plausibility to the possibly overdemocratic 
procedure of computing the variance-weighted averages. 
The result that $\tog > \tgg$ is logically inevitable, but the standard 
Einstein-deSitter model does not satisfy this requirement unless $\hg < 0.55$.
The reference for each measurement is given under the x axis.
The age of the sun is accurately known and is included for reference.
Error bars indicate the reported $1\sigma$ limits.}
\label{fig:figtwo}
\end{figure*}
%%%%%%%%%%%%%%%%%%%%%%%%%%%%%%%%%%%%%%%%%%%%%%%%%

\clearpage

%
%%%%%%%%%%%%%%%%%%%%   Table 1:  Non CMB constraints   %%%%%%%%%%%%%%%%%%%%%%%%%%%%%%%%%%%%%%%%%%%%%%%%%%%%%%%
%{\scriptsize
{\small
\begin{table}[!h]
{\sf
\begin{center}
%\caption{ Non-CMB Constraints$^{a}$}%% \\ [+2.0mm]
\caption{ {\sf Parameter estimates from non-CMB measurements.
I refer to these as constraints. %}%% \\ [+2.0mm]
%\tiny
%\scriptsize
%\noindent
I use the error bars cited here as $1 \sigma$ errors in the likelihood analysis.
The first four constraints are plotted in Fig. 3 B through E.}}
\medskip 
\begin{tabular}{l l l l} \hline
\multicolumn{1}{l}{Method} &
\multicolumn{1}{c}{ Reference} &
\multicolumn{2}{c}{Estimate}\\
\hline
SNe                           &({\it 35})     &$\omzerog = -0.28 \pm 0.16                    $&$ \omfg = 0.27 \pm 0.14                  $\\%&$ H_{o}\t =0.93 \pm 0.06$\\
Cluster mass-to-light         & ({\it 6})     &$\omzerog =  0.19 \pm  0.14                   $&$                                       $\\%&$  $\\
Cluster abundance evolution   &  ({\it 7})   &$\omzerog = 0.17^{+0.28}_{-0.10}              $&$ \omfg = 0.22^{+0.25}_{-0.10}           $\\%&$  $\\
Double radio sources          & ({\it 8})    &$\omzerog = -0.25^{+0.70}_{-0.50}             $&$ \omfg = 0.1^{+0.50}_{-0.20}            $\\%&$  $\\
Baryons                       &({\it 19})    &$\omg\hg^{2/3} = 0.19 \pm 0.12             $&$                                       $\\%&$  $\\
Hubble                        &({\it 16})    &$\hg = 0.68 \pm 0.10                      $&$                                       $\\%&$  $\\
\hline
\end{tabular}
\end{center}
}   %end \sf
\end{table}
}   %end scriptsize or small
\normalsize
%%%%%%%%%%%%%%%%%%%%%%%%%%%%%%%%%%%%%%%%%%%%%%%%%%%%%%%%%%

Here I have combined recent independent measurements of CMB anisotropies ({\it 9}), type Ia SNe ({\it 4, 5}), cluster mass-to-light 
ratios ({\it 6}), cluster abundance evolution ({\it 7}), cluster baryonic fractions ({\it 14}), 
deuterium-to-hydrogen ratios in quasar spectra ({\it 15}),
double-lobed radio sources ({\it 8}), and the Hubble constant ({\it 16})
to determine the age of the universe. % ({\it 12}).
The big picture from the analysis done here is as follows (Figs. 1 and 2):
The Big Bang occurred at $\sim$13.4 Ga. About 1.2 billion years (Gy) later, the halo of
our Galaxy (and presumably the halo of other galaxies)
formed. About 3.5 Gy later, the disk of our Galaxy (and presumably the disks
of other spiral galaxies) formed.
This picture agrees with what we know about galaxy formation.
Even the recent indications of the existence of old galaxies at high redshift ({\it 17})
%(e.g. Yoshii \etal 1998) 
fit into the time framework determined here.
In this sense, the result is not surprising.
What is new is the support given to such a young age by such a wide array of
recent independent measurements.

%\subsection{Method}
\section*{ Method}

Any measurement of a function of $\h,\om$, and $\ol$ can be
included in a joint likelihood
\be
{\mathcal L}(\h, \om, \ol) = \prod^{N}_{i=1} {\mathcal L}_{i}
\ee
which I take as the product of seven of the most recent independent cosmological constraints
(Table 1 and Fig. 3).
%(CMB + six others).
For example, one of the ${\mathcal L}_{i}$ in Eq. 1 represents
the constraints on $\h$.
Recent measurements  can be summarized 
as $\overline{\h} = 0.68 \pm 0.10$ ({\it 16}). 
I represent these measurements
in Eq. 1 by the likelihood,% ({\it 21})
\be
{\mathcal L}_{Hubble}(\h) = \exp{
\left[-0.5 \left( \frac{\h - \overline{\h}}{0.10}\right)^{2}\right]
}
\ee
Another ${\mathcal L}_{i}$ in Eq. 1 comes from measurements of
the fraction of normal baryonic matter in clusters of galaxies ({\it 14})
% (Evrard 1997) 
and estimates of the density of normal baryonic matter in the 

\clearpage

%\vspace{15cm}
\begin{figure*}[!h]
\centerline{\psfig{figure=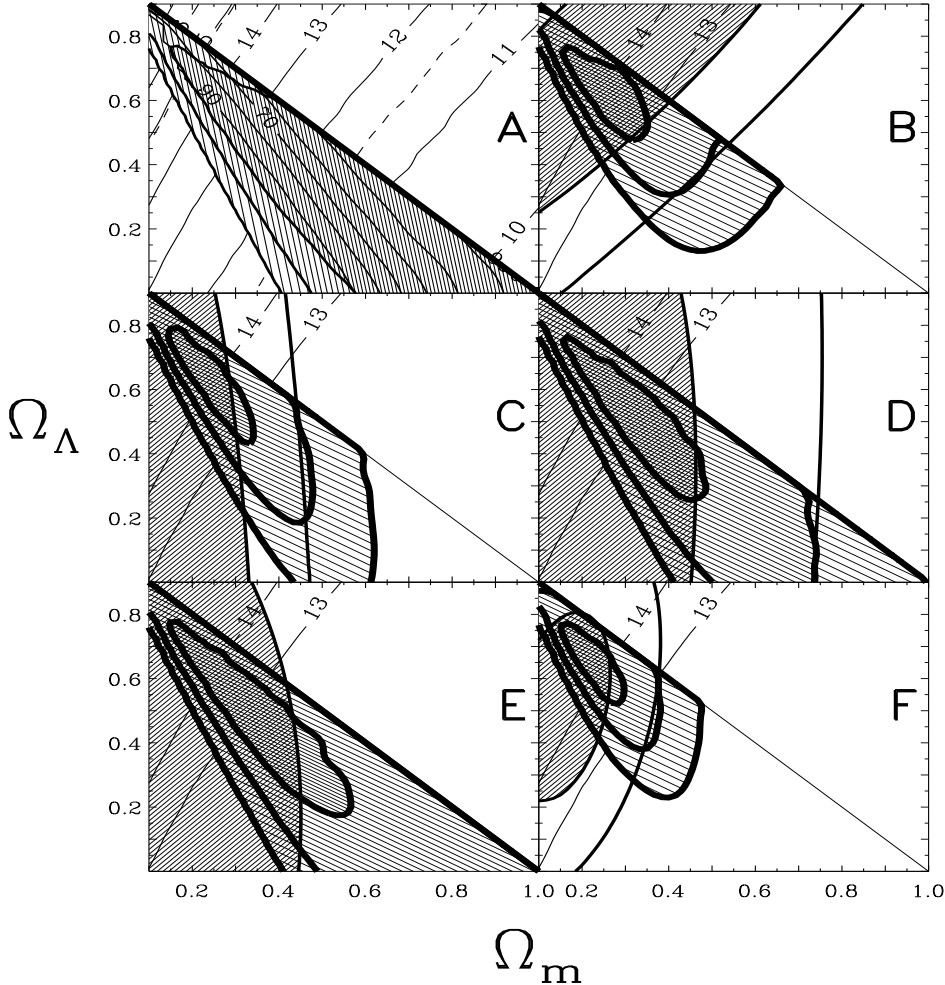,height=12.8cm,width=14.0cm}}
%\centerline{\psfig{figure=~/lambda/figures/plotchi_triangle.ps,height=18.0cm,width=18.cm,bbllx=10pt,bblly=110pt,bburx=594pt,bbury=690pt}}
%BoundingBox: 28 70 594 636
\caption[figure3]{{\footnotesize The regions of the $(\omg, \olg)$ plane preferred by
various constraints. 
({\bf \sf A}) Cosmic microwave background, ({\bf \sf B}) SNe, 
({\bf \sf C}) cluster mass-to-light ratios, ({\bf \sf  D}) cluster abundance evolution, 
({\bf \sf E}) double radio lobes,  and ({\bf \sf F}) all combined.
The power of combining CMB constraints with each of the other constraints (Table 1) is also shown.
The elongated triangles (from upper left to lower right) in ({\bf \sf  A}) are the approximate 
1$\sigma$, 2$\sigma$ and 3$\sigma$ confidence levels of the likelihood from CMB data, ${\mathsf {\mathcal L}_{CMB}}$ ({\it 9}).
%, calculated in Lineweaver (1998).
%
({\bf {\sf A}}) also shows the important $\hg$ dependence of ${\mathsf {\mathcal L}_{CMB}}$. 
The contours within the dark shaded region are of $\hg$ values that maximize 
${\mathsf {\mathcal L}_{CMB}}$ for a given
$(\omg, \olg)$ pair ($\hg = {0.70, 0.90}$ contours are labeled).
This correlation between preferred $\hg$ and preferred $(\omg, \olg)$ helps
${\mathsf {\mathcal L}_{CMB}}(\hg, \omg, \olg)$ constrain $\t$.
In ({\bf \sf B}) through ({\bf \sf E}), thin contours enclose the $ 1 \:\sigma$ (shaded) and $2 \sigma$ 
confidence regions
from separate constraints, and thick contours indicate the  
1$\sigma$, 2$\sigma$ and 3$\sigma$ regions of the combination of  ${\mathsf {\mathcal L}_{CMB}}$ with these same constraints.
%
%The SNe contours ('b') are a combination of the
%Perlmutter \etal (1998) and Riess \etal (1998) results.
%The contours in panel 'c' are obtained from cluster mass-to-light ratios  (Carlberg \etal 1997).
%Panel 'd' shows cluster evolution contours from Bahcall \& Fan (1998).
%Contours in panel 'e' are based on Guerra \etal (1998).
%
%
({\bf{\sf  F}}) shows the region preferred by
the combination of the separate constraints shown in ({\bf \sf B}) through ({\bf \sf E}) (thin contours) as well as the combination 
of ({\bf \sf A}) through ({\bf \sf E})  (thick contours).  
The best fit values are $\olg = 0.65 \pm 0.13$ and $\omg = 0.23 \pm 0.08$.
In ({\bf{\sf  A}}), the thin iso-$\t$ contours (labeled ``10'' through ``14'') indicate the age in Ga 
when $\hg = 0.68$ is assumed.
For reference, the 13- and 14-Ga contours are in all the panels.
To give an idea of the sensitivity of the $\hg$ dependence of these contours, the two additional dashed
contours in ({\bf{\sf  A}}) show the 13-Ga contours for $\hg=0.58$ and $\hg = 0.78$ (the $1 \sigma$ limits of
the principle $\hg$ estimate used in this paper).
In ({\bf{\sf  F}}), it appears that the best fit has $\tog \approx 14.5$ Ga, but 
all constraints shown here are independent of information about $\hg$;
they do not include the $\hg$ dependence of ${\mathsf {\mathcal L}_{CMB}}$, ${\mathsf {\mathcal L}_{baryons}}$ 
or ${\mathsf {\mathcal L}_{Hubble}}$ (Table 1).
}}
\label{fig:figthree}
\end{figure*}
%%%%%%%%%%%%%%%%%%%%%%%%%%%%%%%%%%%%%%%%%%%%%%%%%
\clearpage

\noindent universe [$\obh = 0.015 \pm 0.005$
({\it 15, 18})].
%(see Burles \& Tytler 1998 and Fukugita \etal 1998).
When combined, these measurements yield
$\overline{\om \h^{2/3}} = 0.19 \pm 0.12$ ({\it 19}), which contributes to the likelihood through
\be
{\mathcal L}_{baryons}(\h,\om)= \exp{\left[ -0.5 \left(\frac
{ \om \h^{2/3} - \overline{\om \h^{2/3}}  }{  0.12  } \right)^{2} \right]}
\ee

The $(\om, \ol)$-dependencies of the remaining five constraints  are plotted in Fig. 3 (20).
The 68\% confidence level regions derived from CMB and SNe (Fig. 3, A and B)
are nearly orthogonal, and the region of overlap is relatively small.
Similar complementarity exists between the
CMB and the other data sets (Figs. 3, C through E).
The combination of them all (Fig. 3F) yields 
$\ol = 0.65 \pm 0.13$ and $\om = 0.23 \pm 0.08$ ({\it 21}).

This complementarity is even more important 
(but more difficult to visualize) in three-dimensional parameter space: $(\h,\om, \ol)$.
Although the CMB alone cannot tightly constrain any of these parameters,
it does have a strong preference in the three-dimensional space $(h, \om, \ol)$.
In Eq. 1, I used ${\mathcal L}_{CMB}(\h,\om,\ol)$, which is a generalization of
${\mathcal L}_{CMB}(\om,\ol)$ (Fig. 3A )({\it 22}).
To convert the three-dimensional likelihood ${\mathcal L}(\h, \om, \ol)$ of Eq. 1 into an estimate of the 
age of the universe and into a more easily 
visualized two-dimensional likelihood, ${\mathcal L}(\h,\t)$, I computed the dynamic age 
corresponding to each point in the three-dimensional space $(\h,\om,\ol)$. 
For a given $\h$ and $\t$, I then set ${\mathcal L}(\h,\t)$ equal 
to the maximum value of ${\mathcal L}(\h, \om, \ol)$

\be
{\mathcal L}(\h,\t) = \max \left[ {\mathcal L}(\h,\om,\ol)_{|t(\h,\om,\ol) \approx \t}  \right]
\ee

This has the advantage of explicitly displaying the $\h$ dependence of the $\t$ result.
The joint likelihood ${\mathcal L}(\h,\t)$ of Eq. 4 
yields an age for the universe: $\t = 13.4 \pm 1.6$ Ga (Fig. 4).
This result 
is a billion years younger than
other recent age estimates.

What one uses for ${\mathcal L}_{Hubble}(\h)$ in Eq. 1 is particularly important
because, in general, we expect the higher $\h$ values to yield younger ages.
Table 2 contains results 
from a variety of $\h$ estimates, assuming various central values and various
uncertainties around these values.
The main result $\t = 13.4 \pm 1.6$ Ga has used $\h = 0.68 \pm 0.10$ but
does not depend strongly on the central value assumed for Hubble's constant
(as long as this central value is in the most accepted range, $ 0.64 \le \h \le 0.72$)
or on the uncertainty of $\h$ (unless this uncertainty is taken to be very small).
Assuming an uncertainty of $0.10$, age estimates from using $\h = 0.64,0.68$ and $0.72$
are $13.5, 13.4$ and $13.3$ Ga, respectively (Fig. 2). 
Using a larger uncertainty of $0.15$ with the same $\h$ values
does not substantially change the results, which are ${13.4, 13.3, 13.2}$ Ga, respectively.
For both groups, the age difference is only 0.2 Gy.
If $\t \propto 1/\h$ were adhered to, this age difference would be 1.6 Gy.
Outside the most accepted range the $\h$ dependence becomes stronger and approaches
$\t \propto 1/\h$ ({\it 23}). 

To show how each constraint contributes to the result,
I convolved each constraint separately with Eq. 2 (Fig. 5).
The result does not depend strongly on any one of the constraints (see ``all - x'' results in Table 2).
For example, the age, independent of the SNe data, is 
$\t(all - SNe) = 13.3^{+1.7}_{-1.8}$ Ga,  which differs negligibly from the main result.
The age,  
independent of the SNe and CMB data, is 
$\t(all - CMB -SNe) = 12.6^{+3.4}_{-2.0}$ Ga, which is somewhat lower than the main result but within 
the error bars.
%

%%%%%%%%%%%%%%%%%%%%%%%%%%%%%%%%%%%%%%%%%%%%%%%%
\section*{ The Oldest Objects in Our Galaxy}

The universe cannot be younger than the oldest objects in it.
Thus, 
estimates of the age of the oldest objects in our 
Galaxy are lower limits to the age of the universe (Table 2 and Fig. 2).
A standard but simplified scenario for the origin of our Galaxy has a halo of 
globular clusters forming first, followed by the formation of the Galactic disk.
The most recent measurements of the age of the oldest objects in the Galactic
disk give 
$\td= 8.7 \pm 0.4$ Ga (Table 2).
The most recent measurements of the age 
of the oldest objects in the halo of our Galaxy give
$\tg = 12.2 \pm 0.5$ Ga (Table 2).
The individual measurements are in good agreement with these averages. There are no large outliers.
In contrast to the $\t(\h,\om,\ol)$ estimates obtained above,
all of these age estimates are direct in the sense that they 
have no dependence on a Big Bang model. 

How old was the universe when our Galaxy formed?
If we write this as $\tg + \Delta t  = \t$, then
what is the amount of time ($\Delta t$) between the formation of
our Galaxy and the formation of the universe?
If we had an estimate of $\Delta t$, then we would have an independent estimate
of $\t$ to compare to $\t = 13.4 \pm 1.6$ Ga, obtained above.
However, we have very poor constraints on $\Delta t$.
The simple but plausible estimate $\Delta t \approx 1$ Gy
is often invoked, but estimates range from $\sim 0.1$ to $\sim 5$ Gy,
%$^{Peacock, 20,21,22}$  
and may be even larger ({\it 24, 25}).
%See refs. 19 and 20 
%Stetson \etal 1998, Oswalt \etal 1996  
%and references therein for a review of this issue.
%
This uncertainty in $\Delta t$ undermines the ability of estimates of
the age of the oldest objects in our Galaxy to tell us the age of the universe.
Without $\Delta t$, we cannot infer $\t$ from $\tg$.
The best estimate of $\Delta t$ may come from the difference between 
the age reported here and the estimate of the age of our Galaxy (Table 2). 
Thus $\Delta t = \t - \tg = 13.4 - 12.2 = 1.2 \pm 1.8$ Gy.

The age measurements in Table 2 also indicate that there is a 3.5-Gy period 
between halo and disk formation $(\tg - \td$).
If our Milky Way is typical, then this may be true of other spiral galaxies.
With the best fit values obtained here for the three parameters, 
$(\h, \om, \ol) = (0.72 \pm 0.09, 0.23 \pm 0.08, \: 0.65 \pm 0.13 )$,
the ages $\td$ and $\tg$ can be converted into the redshifts at which the 
disk and halo formed: $z_{disk} = 1.3^{+1.5}_{-0.5}$ and $z_{Gal} = 6.0^{+\infty}_{-4.3}$.
Thus, a diskless epoch should be centered at a redshift between $z_{disk}$ and $z_{Gal}$
$(1.3 \lsim z_{diskless} \lsim  6.0)$.
We would expect fewer disks in the halolike progenitors 
of spiral galaxies in this redshift range.
Studies of galaxy types in the Hubble Deep Field indicate that
this may be the case ({\it 26}).

The requirement  that the universe be older than our Galaxy,
$\t > \tg$, is a consistency test of the Big Bang model. 
The best fit model obtained here passes this test.
There is no age crisis.
This is true even if the high values of $\h$ ($\sim 0.80$)  are correct.
Only at $\h \approx 0.85$ is $\t \approx \tg$.
This consistency provides further support for the Big Bang model,
which the standard model
($\om = 1, \ol = 0$) is unable to match unless 
$\h < 0.55$.

\clearpage
%\vspace{15cm}
\begin{figure*}[!b]
\centerline{\psfig{figure=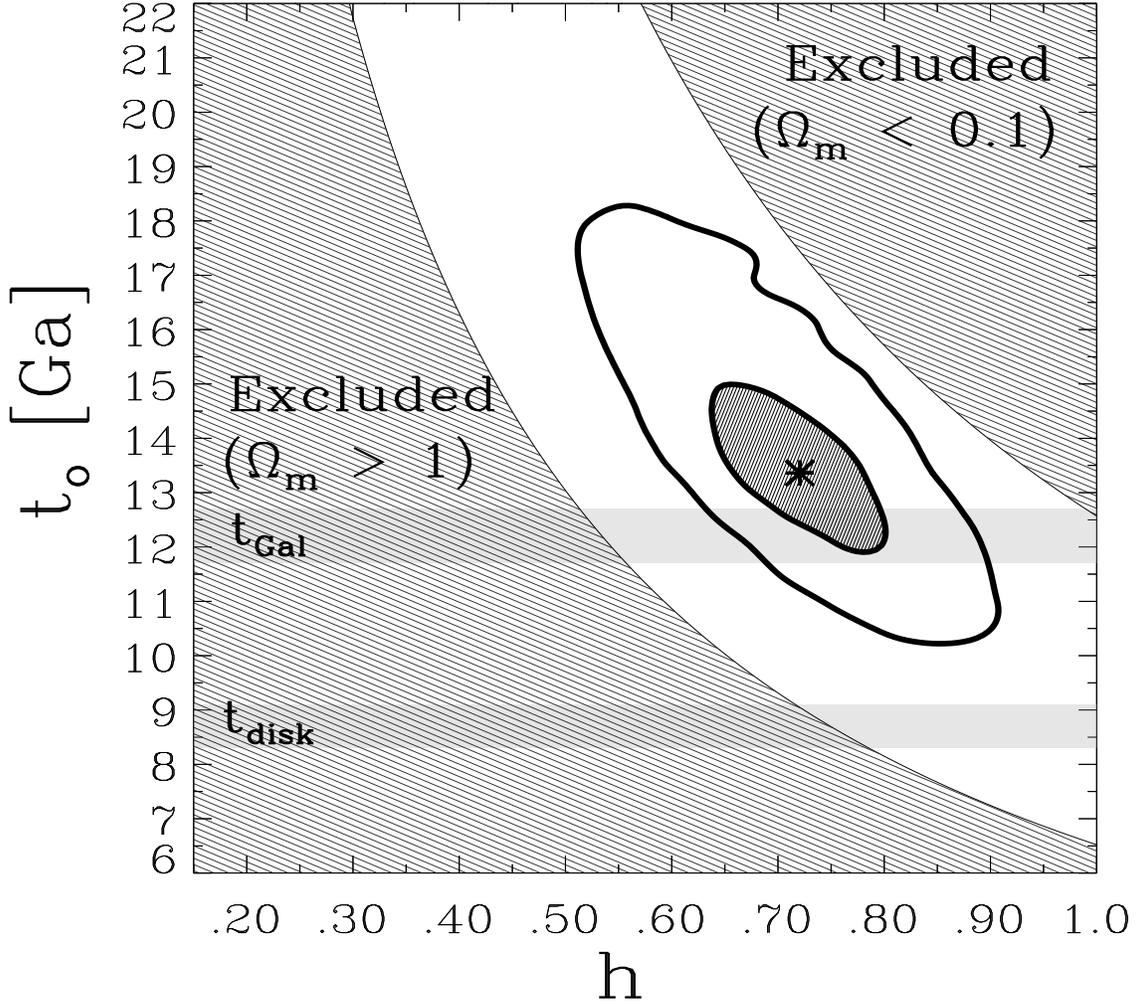,height=15.0cm,width=15.cm}}
%\centerline{\psfig{figure=~/lambda/figures/plotchi_triangle.ps,height=18.0cm,width=18.cm,bbllx=10pt,bblly=110pt,bburx=594pt,bbury=690pt}}
%BoundingBox: 28 70 594 636
\caption{This plot shows the region of the  $\hg - \tog$ plane preferred by the combination of 
all seven constraints.
The result, $\tog = 13.4 \pm 1.6$ Ga, is the main result of this paper.
The thick contours around the best fit (indicated by a star) are at likelihood levels defined by
 ${\mathsf 
{\mathcal L}/{\mathcal L}_{max} = 0.607,
}$ 
and  
${\mathsf 0.135}$,
which approximate  $68\%$ and $95\%$ confidence levels, respectively.
These contours can be projected onto the $\tog$ axis to yield the age result.
This age result is robust to variations in the Hubble constraint
as indicated in Table 2. 
The areas marked ``Excluded'' (here and in Fig. 5) result from the 
range of parameters considered:
$ 0.1 \le \omg \le 1.0$, $\: 0 \le \olg \le 0.9$ with $\omg + \olg \le 1$.
Thus, the upper (high $\tog$) boundary is defined by $(\omg, \olg) = (0.1,0.9)$, and the lower boundary
is the standard Einstein-deSitter model
defined by $(\omg, \olg) = (1, 0)$. Both of these boundary models are plotted in Fig. 1.
The estimates from Table 2 of the age of our Galactic halo ($\tgg$) and the age of
the Milky Way ($\tdg$) are shaded grey. The universe is about
1 billion years older than our Galactic halo.
The combined constraints also yield a best fit value of the Hubble constant 
which can be read off of the x axis ($\hg = 0.73 \pm 0.09$, a
slightly higher and tighter estimate than the input $\hg = 0.68 \pm 0.10$).}
\label{fig:figfour}
\end{figure*}
%%%%%%%%%%%%%%%%%%%%%%%%%%%%%%%%%%%%%%%%%%%%%%%%%

\clearpage
%\vspace{15cm}
\begin{figure*}[!b]
\centerline{\psfig{figure=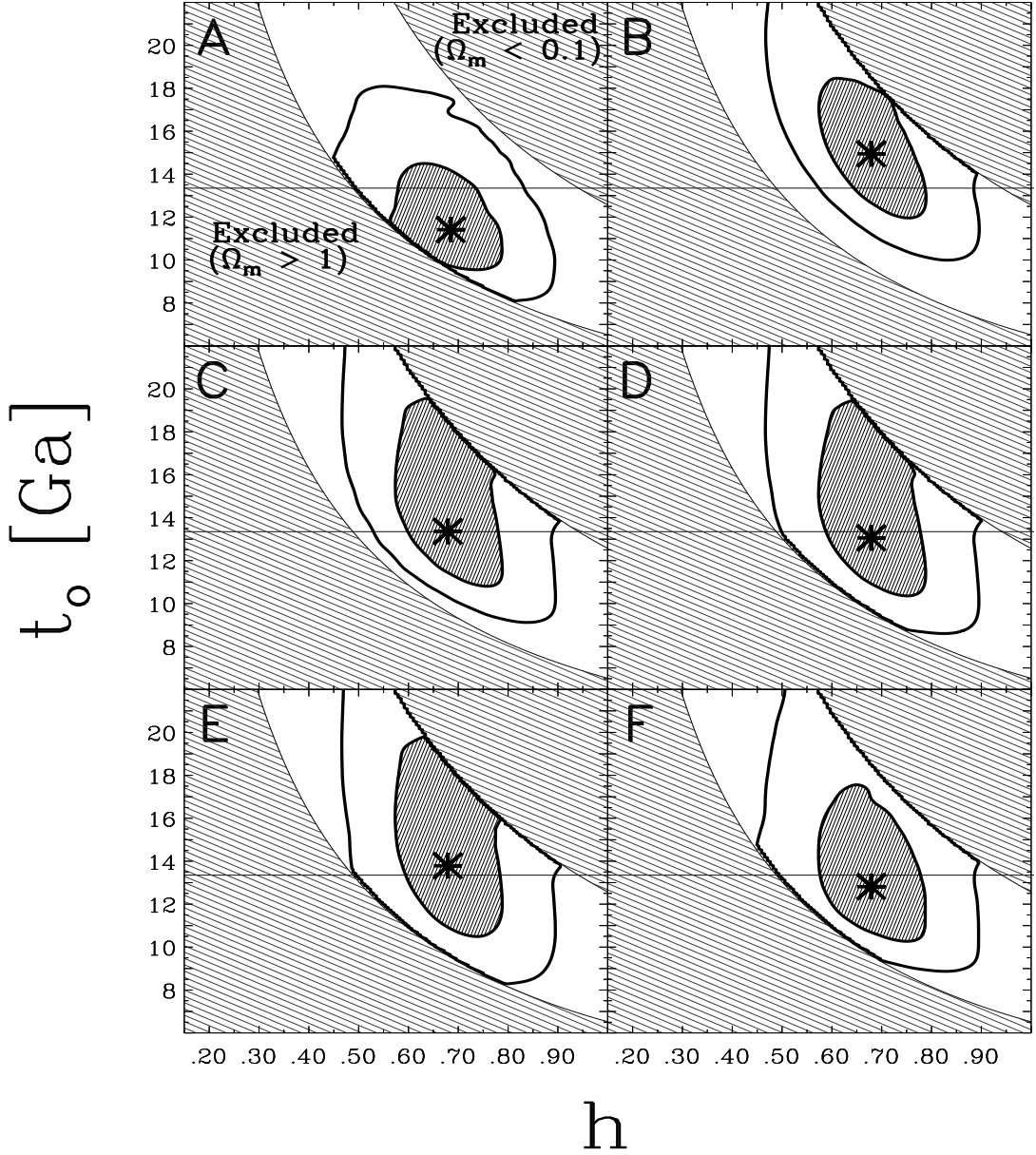,height=16.0cm,width=16.cm}}
%\centerline{\psfig{figure=~/lambda/figures/plotchi_triangle.ps,height=18.0cm,width=18.cm,bbllx=10pt,bblly=110pt,bburx=594pt,bbury=690pt}}
%BoundingBox: 28 70 594 636
\caption{The purpose of this figure is to show how Fig. 4 is built up from 
the seven independent constraints used in the analysis.
All six panels are analogous to Fig. 4 but contain only 
the Hubble constraint [$\hg = 0.68 \pm 0.10$, (Eq. 2)]
convolved with a single constraint:
({\bf \sf A}) cosmic microwave background, ({\bf \sf B}) SNe, 
({\bf \sf C}) cluster mass-to-light ratios, ({\bf \sf  D}) cluster abundance evolution, 
({\bf \sf E}) double radio lobes, and ({\bf \sf F}) baryons (Table 1).
The relative position of the best fit (indicated by a star) and the 13.4-Ga line indicates 
how each constraint contributes to the result.
}
\label{fig:figfive}
\end{figure*}
%%%%%%%%%%%%%%%%%%%%%%%%%%%%%%%%%%%%%%%%%%%%%%%%%

\clearpage

%%%%%%%%%%%%%%%%%%%%%%%%%%%%%%%%%%%%%%%%%%%%%
\section*{ Comparison with Previous Work}

The goal of this paper is to determine the absolute age of the universe $\t(\h,\om,\ol)$.
Knowledge of $\h$ alone cannot be used to determine $\t$ with much accuracy. 
For example, the estimate $\h = 0.68 \pm 0.10$ corresponds to $8 < \t < 22$ Ga
(Fig. 4).
Similary, knowledge of $(\om,\ol)$ yields $H_{o}\t (\om,\ol)$ not $\t$
($H_{o}$ is the usual Hubble constant).
When one inserts a preferred value of $\h$ into a $H_{o}\t$ result, one is not taking into consideration
the correlations between preferred $\h$ values and preferred $(\om,\ol)$ values
that are inherent, for example, in ${\mathcal L}_{CMB}(\h, \om, \ol)$ and ${\mathcal L}_{baryons}(\h, \om)$. 
The preferred values of $\h$ in these likelihoods depend on $\om$ and $\ol$.
Perlmutter \etal ({\it 4}) used SNe measurements to constrain $(\om, \ol)$ and 
obtained values for $H_{o}\t$. To obtain $\t$, they did the analysis with $\h$
set equal to the value preferred by their SNe data, $\h = 0.63$.
Their result is $\t = 14.5\pm 1.0 (0.63/h)$ Ga. 
When a flat universe is assumed, they obtain $\t^{flat} = 14.9^{+1.4}_{-1.1} (0.63/h)$ Ga.
Riess \etal ({\it 5}) found $\h = 0.65 \pm 0.02$ from their SNe data.
Marginalizing over this Hubble value and over $\ol$ and $\om$,
they report $\t= 14.2 \pm 1.7 $ Ga.
When a flat universe is assumed, their results yield $\t^{flat} = 15.2 \pm 1.7$ Ga.
The Perlmutter \etal ({\it 4}) and Riess ({\it 5}) results are 
in good agreement.
When I assume $\h = 0.64 \pm 0.02$, I get
$\t = 14.6 ^{+1.6}_{-1.1}$ Ga.
This result is plotted in Fig. 2 to illustrate the important influence on the result of using a small 
$\h$ uncertainty. 
Efstathiou \etal ({\it 12}), on the basis of a combination of CMB and Perlmutter \etal ({\it 4}) SNe data,  have 
estimated $\t = 14.6 (\h/0.65)^{-1}$ Ga.
I used  $\h = 0.65 \pm 0.0$ with this data combination to get
$\t= 14.5^{+1.2}_{-1.0}$ Ga. However, when I used $\h = 0.65 \pm 0.10$, 
the result is 0.7 Gy lower ($\t = 13.8^{+3.2}_{-1.4}$ Ga). 
To obtain the main result, I used uncertainties large enough to reflect our knowledge of $\h$
on the basis of many sources. The use of a larger $\h$ uncertainty contributes to 
the substantially younger ages found here ({\it 23}).

A potential problem with the SNe ages is the high region, $(\om, \ol) \approx (0.8,1.5)$, which 
dominates the SNe fit.  
This region is strongly disfavored by the six other constraints considered here (see Fig. 3).
These high $(\om, \ol)$ values allow lower ages than the $\tf$ SNe results because the 
slope of the iso-$\t$ contours (Fig. 3B) is larger than the slope of the SNe contours.
The $\tf$ results are not as subject to this problem and are the results most analogous to the
result reported here,  despite the fact that the SNe $\tf$ results are less consistent with 
the result reported here.
There are several independent cosmological measurements which have not been
included in this analysis either because 
a consensus has not yet been reached 
[gravitational lensing limits ({\it 27-30})]
or because the analysis of the measurements has not been done in a way
that is sufficiently free of conditioning on certain parameters [local velocity field limits ({\it 31})].
Doubts about some of the observations used here are 
discussed in ({\it 32}).
% Dekel, Dressler \& White (1997).
%
There has been speculation recently that the evidence for $\ol$ is really evidence
for some form of stranger dark energy that we have incorrectly been interpreting as $\ol$.
Several workers have tested this idea.
The evidence so far indicates that the cosmological constant interpretation
fits the data as well as or better than an explanation based on 
more mysterious dark energy ({\it 4, 33, 34}).
% (Garnavich \etal 1998, Perlmutter \etal 1999, 1998).

\newpage
%%%%%%%%%%%%%%%%%%%%%%%%%%%%%%%  Age table  %%%%%%%%%%%%%%%%%%%%%%%%%%%%%%%%%%%%%%%%%%%%%
%{\scriptsize
%{\tiny
{\footnotesize
\begin{table}
{\sf 
\begin{center}
%\caption{ Age Estimates of Our Galaxy and Universe}%% \\ [+2.0mm]
\caption{ {\sf {\footnotesize
Age estimates of our Galaxy and universe ({\it 36}).
``Technique'' refers to the method used to make the age estimate. 
OC, open clusters; WD, white dwarfs; LF, luminosity function; GC, globular clusters;
M/L, mass-to-light ratio; and cl evol, cluster abundance evolution.
The averages are inverse variance-weighted averages of the individual measurements. The sun is not included in the
disk average.
``Isotopes'' refers to the use of relative isotopic abundances of long-lived species as indicated by absorption
lines in spectra of old disk stars.
The ``stellar ages'' technique uses main sequence fitting and the new Hipparcos subdwarf calibration.
%$^{e}$ inverse variance weighted average of the halo age measurements.\\
%Ratra \etal 1998 cite $12.5 \pm 2.5$ Gyr\\
%The bold face age estimates are plotted in Fig. 2.
``All'' means that all six constraints in Table 1 and the CMB constraints were used in Eq. 1.
``All-x'' means that all seven constraints except constraint x were used in Eq. 1.
Figures. 3 and 5 and the all - x results indicate a high level of agreement
between constraints and the lack of dependence on any single constraint.
Thus, there is a broad consistency between the ages preferred by the CMB and the six other independent constraints. Figure 2 presents all of the disk and halo age estimates.
}}}

\medskip

\begin{tabular}{l c l l l} \hline
\multicolumn{1}{l}{Technique} &
\multicolumn{1}{c}{Reference} &
\multicolumn{1}{l}{$\hg$ Assumptions} &
\multicolumn{1}{l}{Age (Ga) } &
\multicolumn{1}{l}{Object}\\
\hline
%Guenther \etal (1997)
Isotopes           &  ({\it 37})     &         None          &$4.53 \pm 0.04              $& Sun  \\
\hline
%                   &                 &                       &                             &           \\
%Chaboyer \etal (1999)
Stellar ages       &  ({\it 38})     &         None          &$ 8.0 \pm 0.5               $&  Disk OC \\
%Leggett \etal (1998) 
WD LF              &  ({\it 39})     &         None          &$ 8.0 \pm 1.5               $&  Disk WD \\
%Carraro \etal (1999)
Stellar ages       &  ({\it 40})     &         None          &$ 9.0 \pm 1                 $&  Disk OC \\
%Oswalt \etal(1998) 
WD LF              &  ({\it 25})     &         None          &$ 9.7^{+0.9}_{-0.8}         $&  Disk WD \\
%Phelps (1997)
Stellar ages       &  ({\it 41})     &         None          &$ 12.0^{+1.0}_{-2.0}        $&  Disk OC \\
                   &                 &         None          &$ 8.7 \pm 0.4               $&  $\tdg$(avg) \\
%                   &                 &                       &                             &           \\
\hline
%Chaboyer \etal(1998)
Stellar ages       & ({\it 42})      &         None          &$ 11.5 \pm 1.3              $&  Halo GC\\
%Gratton \etal(1997)
Stellar ages       & ({\it 43})      &         None          &$ 11.8^{+1.1}_{-1.3}        $&  Halo GC\\
%Reid(1997)                   
Stellar ages       & ({\it 44})      &         None          &$ 12 \pm 1                  $&  Halo GC\\
%Salaris \etal(1997)
Stellar ages       & ({\it 45})      &         None          &$ 12 \pm 1                  $&  Halo GC\\
%Grundahl \etal(1998)
Stellar ages       & ({\it 46})      &        None           &$ 12.5 \pm 1.5              $&  Halo GC\\
%Cowan \etal (1998)
Isotopes           & ({\it 47})      &        None           &$ 13.0 \pm 5                $&  Halo stars\\
%Jimenez \etal(1998)
Stellar ages       & ({\it 48})      &        None           &$ 13.5 \pm 2                $&  Halo GC\\
%Pont \etal(1998)
Stellar ages       & ({\it 49})      &        None           &$ 14.0^{+2.3}_{-1.6}        $&  Halo GC\\
                   &                 &        None           &$ 12.2 \pm 0.5              $& $\tgg$ (avg)  \\
\hline
%
%Perlmutter \etal                   &  
SNe                & ({\it 4})      & $0.63  \pm 0.0$    &$  14.5 \pm 1.0 $& Universe  \\
SNe (flat)         & ({\it 4})      & $0.63  \pm 0.0$    &$  14.9^{+1.4}_{-1.1}*$& Universe \\
% Reiss  \etal (4)                       &  
SNe                & ({\it 5})      & $0.65  \pm 0.02$   &$  14.2 \pm 1.7 $& Universe \\
SNe (flat)         & ({\it 5})      & $0.65  \pm 0.02$   &$ 15.2 \pm 1.7 * $& Universe \\
%\hline
% this work                         &  
All                & This work    & $0.60  \pm 0.10$     &$  15.5^{+2.3}_{-2.8} $& Universe   \\
All              & This work          & $0.64  \pm 0.10$     &$13.5^{+3.5}_{-2.2}* $& Universe  \\
   All              & This work          & $0.68  \pm 0.10$     &$13.4^{+1.6}_{-1.6}* $& Universe  \\
All                &  This work         & $0.72  \pm 0.10$     &$13.3^{+1.2}_{-1.9}* $& Universe  \\
  All              &   This work         & $0.76  \pm 0.10$     &$  12.3^{+1.9}_{-1.6} $& Universe  \\
  All              &  This work          & $0.80  \pm 0.10$     &$  11.9^{+1.9}_{-1.6} $& Universe  \\
% the following three have been cut to reduce the ``concentration on h'' as requested by one of the referees.
%%%   --                              &    --           & $\h=0.64  \pm 0.15$  &$  13.4^{+3.4}_{-1.9} $& --  \\
%%%   --                              &    --           & $\h=0.68  \pm 0.15$  &$  13.3^{+1.5}_{-2.1} $& --  \\
%%%   --                              &    --           & $\h=0.72  \pm 0.15$  &$  13.2^{+1.4}_{-2.7} $& --  \\
%
  All              &   This work          & $0.64  \pm 0.02$  &$14.6^{+1.6}_{-1.1}* $& Universe  \\
  All$-$CMB        &   This work           & $0.68  \pm 0.10$     &$  14.0^{+3.0}_{-2.2}  $& Universe \\
  All$-$SNe        &    This work          & $0.68  \pm 0.10$     &$  13.3^{+1.7}_{-1.8}  $& Universe \\
  All$-$M/L        &    This work          & $0.68  \pm 0.10$     &$  13.3^{+1.9}_{-1.7}  $& Universe \\
  All$-$cl evol    &    This work          &  $0.68  \pm 0.10$    &$  13.3^{+1.7}_{-1.4}  $& Universe \\
  All$-$radio      &    This work          &  $0.68  \pm 0.10$    &$  13.3^{+1.7}_{-1.5}  $& Universe \\
  All$-$baryons    &    This work          &  $0.68  \pm 0.10$    &$  13.4^{+2.6}_{-1.5}  $& Universe \\
  All$-$Hubble     &    This work          &      None            &$ < 14.2               $& Universe \\
  All$-$CMB$-$SNe  &    This work          &  $0.68  \pm 0.10$    &$  12.6^{+3.4}_{-2.0}  $& Universe \\

\hline

\end{tabular}\\
\end{center}
\indent $*$ Also plotted in Fig. 2.

}  %end \sf
\end{table}
}   %end scriptsize or end small

\normalsize

\clearpage

\renewcommand{\baselinestretch }{1}

\smallskip

%%%%%%%%%%%%%%%%%%%%%%%%%%%%%%%%%%%%%%%%%%%%%%%%%%%%%%%%%%%%%%%%%%%%%%%%%%%
\end{document}